\documentstyle[prb,aps]{revtex}

\title{Temperature Dependence of the Cu(2) NQR
Line Width in YBa$_2$Cu$_3$O$_{7-y}$}

\author{A.\,V.\,Dooglav, M.\,V.\,Eremin, Yu.\,A.\,Sakhratov, A.\,V.\,Savinkov}

\address{Kazan State University, 420008 Kazan, Russia}

\begin{document}

\date{\today }

\maketitle

\begin{abstract}
Systematic measurements of the $^{63}$Cu(2) NQR line width were performed in
underdoped YBa$_2$Cu$_3$O$_{7-y}$ samples over the temperature range 4.2\,K
$<T<300$\,K. It was shown that the copper NQR line width monotonically
increases upon lowering temperature in the below-critical region, resembling
temperature behavior of the superconducting gap. The observed dependence is
explained by the fact that the energy of a condensate of sliding charge-current
states of the charge-density-wave type depends on the phase of order parameter.
Calculations show that this dependence appears only at $T<T_c$. Quantitative
estimates of the line broadening at $T<T_c$ agree with the measurement results.
\\ To be published in JETP Lett. {\bf 74}, No. 2, pp.103-106 (2001).
\end{abstract}

\twocolumn

It is generally believed that the Cu(2) NQR lines in high-$T_c$ superconductors
of the 123 type with oxygen index close to 7 are strongly broadened due to
crystal defects (oxygen vacancies and interstitials, twinning, impurity phases,
lattice imperfections, etc.). This is confirmed by the fact that the copper NQR
line width in the 1248 compound with stoichiometric composition is appreciably
smaller than in the 1237 compound~\cite{bib:A1}. The NQR line widths in
YBa$_2$Cu$_3$O$_{7-y}$ (Y1237) samples were studied by different scientific
groups and found to vary over a rather wide range: 200-350\,kHz at room
temperature. The width virtually does nor change upon changing temperature from
room to critical ($T_c$). However, starting at $T_c$ (or below $T_c$ at small
doping~\cite{bib:A2}), the line width starts to rapidly increase. Although the
reasons for this broadening are as yet unclear, its quadrupolar origin is
evident from the comparison of the line widths of the $^{63}$Cu and $^{65}$Cu
isotopes. As a rule, the larger the NQR line width at $T>T_c$ the larger the
broadening. If one assumes that the line width at $T>T_c$ is due to lattice
defects, then one should accept that these defects influence, at least in part,
the broadening of the NQR lines below $T_c$. There are grounds to believe that
the broadening at $T<T_c$ depends on the doping level of the sample. For
instance, the Cu(2) NQR line width measured in~\cite{bib:A3} for an overdoped
Y1237 sample virtually did not change in the temperature range 300-10\,K,
although the line width (about 290\,kHz) was not too small; i.e., the sample
studied in that work was not defectless.

Therefore, the temperature behavior of the NQR line width of the in-plane
copper in a 123-type superconductor is nontrivial and calls for a better
understanding. In this work, we report the results of a detailed line-width
study for two Y123 samples and analyze the possible reasons of broadening. To
interpret the additional quadrupole broadening below $T_c$, we suggest to take
into account the retarded motion of charge-current waves and substantiate this
by the mean-field calculations.

We carried out experiments using the same two paraffin-packed Y1237 powder
samples with crystallite sizes $\sim 30$\,$\mu m$ as in~\cite{bib:A4}. The
critical temperatures (onset of transition) were 91.6\,K (sample 1) and 91.2\,K
(sample 2). The copper NQR spectra were measured on a broad-band coherent
pulsed NQR/NMR spectrometer over the temperature range 300-4.2\,K. Examples of
the Cu(2) NQR spectra recorded at temperatures 300 and 4.2\,K are given in
Fig.1. In both samples, the $^{63}$Cu(2) NQR line shape is well fitted to the
Lorentzian curve with temperature-dependent FWHH (Fig.2). One can see that in
both samples the line width weakly varies with temperature from room to a
temperature of $\sim 120$\,K. The line begins to broaden below 120\,K and
broadens sharply below $T_c$. Because of a large error in measuring the line
width in sample 1, one cannot judge confidently any details of its behavior
below $T_c$. The line width in sample 2 has a clear maximum at $T=47$\,K and a
minimum at 35\,Ê. These features are due to the spin–spin relaxation processes,
which were discussed in~\cite{bib:A4}.

Before analyzing the NQR line broadening near $T_c$, we adduce several
arguments in favor of the assumption that this effect is intrinsic in nature,
i.e., that it is inherent in the compound as such and is not associated with
the surface effect of crystallites. It is well known that the penetration depth
of an rf magnetic field exciting nuclear spin system decreases upon transition
into the superconducting state, as a result of which the NMR/NQR signal
intensity also decreases, because the nuclei in the crystallite bulk drop out
of the observation. Therefore, the relative contribution to the NMR/NQR signal
from the nuclei located in the surface layers of crystallites increases below
$T_c$. Since the crystal lattice is distorted in the surface layers, the NQR
line of the nuclei from this layer is ordinarily broader than the bulk line.
The thickness of a distorted surface layer is governed by the crystal
properties and is about 0.2\,$\mu m$ at worst (crystals mechanically dry-ground
in a mortar and having no cleavage planes~\cite{bib:A5}). Our calculations have
shown that even if the surface NQR line is broader than the bulk line by a
factor of 10, then, for the crystallites 30\,$\mu m$ in size and a 2.5-fold
decrease in signal intensity below $T_c$ (as is the case in our experiments), a
faintly visible influence of surface on the NQR line broadening occurs for a
surface layer one micrometer thick. Since the relative line broadening in our
experiments is equal to several tens percent, we assume that the surface effect
on the Cu(2) NQR line width can be ignored.

We believe that the above-mentioned features of the temperature behavior of
line width shed new light on the coexistence regime of the pseudogap and
superconducting phases in cuprates. Indeed, the energy of a quasiparticle in a
superconductor in the presence of a homogeneous spatial modulation with wave
vector $\bf Q=(\pi,\pi)$ is given by the expression
\begin{eqnarray}
\label{eq:1}
E^{2}_{1\bf k,2\bf k} = \frac{1}{2} \Bigl( \epsilon^{2}_{\bf k} +
\epsilon^{2}_{\bf k+Q} \Bigr) + |G_{\bf k}|^{2}
+ \nonumber \\
+ \frac{1}{2} \Bigl(|\Delta_{\bf k}|^{2} + |\Delta_{\bf k+Q}|^{2} \Bigr) +
|U_{\bf k}|^{2}
\pm \nonumber \\
\pm  \Biggl\{ \frac{1}{4} \Bigl[ \epsilon^{2}_{\bf k} - \epsilon^{2}_{\bf k+Q}
+ |\Delta_{\bf k}|^{2} - |\Delta_{\bf k+Q}|^{2} \Bigr]^{2}
+ \nonumber \\
+ |G_{\bf k}|^{2} \Bigl[ \Bigl( \epsilon_{\bf k} + \epsilon_{\bf k+Q}
\Bigr)^{2} + |\Delta_{\bf k}|^{2} + |\Delta_{\bf k+Q}|^{2} \Bigr]
- \nonumber \\
- \Delta^{\ast}_{\bf k} \Delta_{\bf k+Q} G_{\bf k}^{2} - \Delta_{\bf k}
\Delta^{\ast}_{\bf k+Q} (G^{\ast}_{\bf k})^{2}
+ \nonumber \\
+ |U_{\bf k}|^{2} \Bigl[ \Bigl( \epsilon_{\bf k} - \epsilon_{\bf k+Q}
\Bigr)^{2} + |\Delta_{\bf k}|^{2} + |\Delta_{\bf k+Q}|^{2} \Bigr]
+ \nonumber \\
+ \Delta_{\bf k} \Delta_{\bf k+Q} (U^{\ast}_{\bf k})^{2} + \Delta^{\ast}_{\bf
k} \Delta^{\ast}_{\bf k+Q} U^{2}_{\bf k}
+ \nonumber \\
+ 2\epsilon_{\bf k} \Bigl[ U^{\ast}_{\bf k} G_{\bf k} \Delta_{\bf k+Q} + U_{\bf
k} G^{\ast}_{\bf k} \Delta^{\ast}_{\bf k+Q} \Bigr]
+ \nonumber \\
+ 2\epsilon_{\bf k+Q} \Bigl[ \Delta_{\bf k} U^{\ast}_{\bf k} G^{\ast}_{\bf k} +
\Delta^{\ast}_{\bf k} U_{\bf k} G_{\bf k} \Bigr] \Biggr\}^{1/2},
\end{eqnarray}
where the Hamiltonian of the model is chosen as in~\cite{bib:A6}, $\Delta_{\bf
k}$ is the order parameter in the superconducting phase; the real part of the
parameter $G_{\bf k}=S_{\bf k}+iD_{\bf k}$ corresponds to the spatially
modulated charge or, in other words, charge density waves (CDWs), and the
imaginary part of this parameter corresponds to the orbital currents
circulating in the neighboring unit cells in opposite directions. The imaginary
part of $G_{\bf k}$ has the $d$-type symmetry, while its real part is
characterized by the $s$-type symmetry~\cite{bib:A6}. In the underdoped
samples, the critical temperature $T^\ast$ of parameter $D$ is higher than the
superconducting transition temperature $T_c$. The parameter $U_{\bf k}$
describes the inhomogeneous state of the superconductor. According
to~\cite{bib:A4}, its value in the crystals under study is comparatively small.

The CDWs and orbital currents are decelerated and, in the limit, pinned by the
lattice potential and defects, because the quasiparticle energy depends on the
phase of order parameter. Let us substitute $G_{\bf k}=|G|e^{i\varphi}$
into~(\ref{eq:1}) and trace how $E_{\bf k}$ depends on the phase~$\varphi$. It
is seen from~(\ref{eq:1}) that the energy of a condensate of orbital currents
and CDWs is independent of~$\varphi$ at $T_{c}<T<T^{\ast}$. The sliding charge–
current states are not decelerated (at least in the mean-field approximation)
and, hence, do not contribute to the broadening of the copper NQR lines. It is
also seen from~(\ref{eq:1}) that the phase dependence appears as $\Delta_{\bf
k}$ becomes nonzero, i.e., at $T=T_c$, and this dependence becomes stronger
(proportional to $\Delta_{\bf k}(T)$ squared) upon lowering the temperature.
Qualitatively, this explains a monotonic increase in the Cu(2) NQR line width
at $T<T_c$. It also becomes clear that the broadening in overdoped samples is
absent~\cite{bib:A3} because $|G|=0$ in them.

For the quantitative estimates, let us consider Fourier component of the
spatially modulated charge on the copper nuclei:
\begin{equation}
\label{eq:2}
e_{Q} = \frac{1}{N} \sum_{j} \delta_{j} \exp (-iQR_{j}),
\end{equation}
where $\delta_j$ is the number of doped holes per a unit cell of the CuO$_2$
bilayer and $N$ is the number of unit cells. The homogeneous part $\delta_0$ of
distribution does not contribute to $e_Q$. By separating the term proportional
to $\Delta_{\bf k}$ in the charge modulation Fourier amplitude, one has,
according to~\cite{bib:A6},
\begin{eqnarray}
\label{eq:3}
e_{Q} \approx \frac{2+\delta_{0}}{8N} \sum_{\bf k}
\frac{\Delta^{2}_{\bf k}(G_{\bf k}+G^{\ast}_{\bf k})} {(E^{2}_{1\bf
k}-E^{2}_{2\bf k})}
\times \nonumber \\
\times \Biggl[ \frac{1}{E_{1\bf k}} \tanh \Biggl( \frac{E_{1\bf k}}{2k_{B}T}
\Biggr) - \frac{1}{E_{2\bf k}} \tanh \Biggl( \frac{E_{2\bf k}}{2k_{B}T} \Biggr)
\Biggr].
\end{eqnarray}
For simplicity, we restrict ourselves to the case $U_{\bf k}=0$. Note that
$e_Q$ is zero if the real component of the pseudogap ($S_{\bf k}$) is absent.
For the numerical estimates, we specify the dispersion relation for
quasiparticles as
\begin{eqnarray}
\label{eq:4} \epsilon_{\bf k} = \frac{2+\delta_{0}}{2} \Bigl[t_{1} ( \cos k_{x}
+ \cos k_{y})
+ \nonumber \\
+ 2 t_{2} \cos k_{x} \cos k_{y} + t_{3} ( \cos 2k_{x} + \cos 2k_{y}) \Bigr],
\end{eqnarray}
where $t_1=78$\,meV, $t_2=0$, and $t_3=12$\,meV~\cite{bib:A6}. In accordance
with the photoemission data~\cite{bib:A7}, we assume that $\Delta_{\bf
k}=\Delta(T)(\cos{k_x}-\cos{k_y})$, $D_{\bf k}=D(T)(\cos{k_x}-\cos{k_y})$, and
$\delta_0 \approx 0.3$. The difference between $|G_{\bf k}|$ and
$|\cos{k_x}-\cos{k_y}|$ for underdoped
Bi$_2$Sr$_2$Ca$_{1-x}$Dy$_x$Cu$_2$O$_{8+y}$  at $T>T_c$ (see Fig.2
in~\cite{bib:A7}) allows the relative value of $S$ component to be estimated as
$S(T)/D(T)\approx 0.05 \div 0.1$. For the order-of-magnitude estimates, we
ignore the dependence of $S_{\bf k}$ on the wave vector; i.e., we assume that
$S_{\bf k}$ depends only on temperature. The electric field gradient at the
copper nuclei is mainly comprised of the lattice and valence
contributions~\cite{bib:A8}. The valence contribution from the copper hole
$d_{x^2-y^2}$ orbital dominates and equals approximately $V_{zz}(val ) \approx
70$\,MHz. In the presence of spatially modulated charge, the copper NQR line
width can be estimated as $\Delta \nu \approx 2V_{zz}(val)e_Q$. The temperature
dependence of line width thus calculated is shown in Fig.2 by the solid line.
Qualitatively, it agrees well with the experimental results. The agreement in
order of magnitude is also observed. It is worth noting that~(\ref{eq:2}) holds
in the limit of CDW motion slow compared to the period of probe field and does
not assume that the CDWs are completely pinned.

A complete picture of the CDW effect on the line width, clearly, should include
a contribution from the component that already exists at $T>T_c$. At
$T<T^\ast$, it is given by the expression
\begin{equation}
\label{eq:5} e_{Q}' \approx \frac{2+\delta_{0}}{4N} \sum_{\bf k} \frac{S_{\bf
k}}{E_{1\bf k} - E_{2\bf k}} \Bigl[ f(E_{1\bf k}) - f(E_{2\bf k}) \Bigr ],
\end{equation}
where
\begin{equation}
\label{eq:6} E_{1\bf k,2\bf k} = \frac{\epsilon_{\bf k} + \epsilon_{\bf
k+Q}}{2} \pm \frac{1}{2} \Bigl[ (\epsilon_{\bf k} - \epsilon_{\bf k+Q})^{2} +
4|G_{\bf k}|^{2} \Bigr]^{1/2},
\end{equation}
and $f(E_{\bf k})$ is the Fermi distribution function.

We did not observe the $e_{Q}'$ component in our experiments; though noticeable
in Fig.2, the line broadening is very small in the region slightly above $T_c$.
Within the framework of the picture suggested, this can be explained by the
averaging of $e_{Q}'$ over the relatively fast CDW motion above $T_c$. In this
respect, an important role is played by the fact that the critical temperature
of parameter $S(T)$ is smeared due to strong fluctuations that are inherent in
the pseudogap phase. As known, these fluctuations are not taken into account in
the mean-field approximation and, thus, require special calculations.

The presence of the $S$ component can also be derived from the photoemission
data for Bi$_2$Sr$_2$Ca$_{1-x}$Dy$_x$Cu$_2$O$_{8+y}$ at $T>T_c$~\cite{bib:A7}.
However, its influence on the photoemission spectra is on the verge of
experimental accuracy and, probably, was not revealed in~\cite{bib:A7} for this
reason. In our case, the $iD$ component does not contribute to the quadrupolar
width, and, hence, the $S$ component is more pronounced.

In summary, the measurements of the copper NQR line width in near-optimum doped
YBa$_2$Cu$_3$O$_{7-y}$ have shown that it starts to monotonically increase at
$T<T_c$, resembling, by its temperature dependence, the behavior of the order
parameter for superconducting gap. We relate the observed dependence to the
presence of $S$-type component in the order parameter of the pseudogap phase.
We used the mean-field approximation to show that the energy of a condensate of
charge-current states depends on the phase of order parameter only at $T<T_c$.
The quantitative estimates agree with the experimental results.

This work was supported by the Russian program ``Superconductivity (project No.
9814-1) and, in part, by the BRHE (grant No. REC-007).

Fig.1. $^{63,65}$Cu(2) NQR spectra of YBa$_2$Cu$_3$O$_{7-y}$ at 300 and 4.2\,K.

Fig.2. Temperature dependence of the $^{63}$Cu(2) NQR line width in
YBa$_2$Cu$_3$O$_{7-y}$. The solid line is the result of calculation (see text)
with the following parameters: $T_c=91$\,K, $T^\ast=120$\,K,
$\Delta(T=0K)=25$\,meV, $D(T=0K)=30$\,meV, $S(T)/D(T)=0.083$.


\begin{thebibliography}{8}

\bibitem{bib:A1}
D. Brinkmann, in {\it Materials and Crystallographic Aspects of
HTc-Superconductivity} (Kluwer Academic Publishers, 1994), pp.225-248.

\bibitem{bib:A2}
S. Kr\"amer and M. Mehring, Phys. Rev. Lett. {\bf 83}, 396 (1999).

\bibitem{bib:A3}
K. Kumagai, K. Nozaki, and Y. Matsuda, Phys. Rev. B. {\bf 63}, 144502 (2001).

\bibitem{bib:A4}
M.\,V. Eremin, Yu.\,A. Sakhratov, A.\,V. Savinkov, {\it et al.}, Pis'ma Zh.
Eksp. Teor. Fiz. {\bf 73}, 609 (2001) [JETP Lett. {\bf 73}, 540 (2001)].

\bibitem{bib:A5}
R.\,Yu. Abdulsabirov, A.\,A. Bukharaev, M.\,R. Zhdanov, {\it et al.},
cond-mat/9808163.

\bibitem{bib:A6}
S.\,V. Varlamov, M.\,V. Eremin, and I.\,M. Eremin, Pis'ma Zh. Eksp. Teor. Fiz.
{\bf 66}, 726 (1997) [JETP Lett. {\bf 66}, 569 (1997)]; M.\,V. Eremin and
I.\,A. Larionov, Pis'ma Zh. Eksp. Teor. Fiz. {\bf 68}, 583 (1998) [JETP Lett.
{\bf 68}, 611 (1998)].

\bibitem{bib:A7}
J.\,M. Harris, Z.-X. Shen, P.\,J. White, {\it et al.}, Phys. Rev. B. {\bf 54},
R15665 (1996).

\bibitem{bib:A8}
C.\,H. Pennington, D.\,J. Durand, C.\,P. Slichter, {\it et al.}, Phys. Rev. B.
{\bf 39}, 2902 (1989).

\end{thebibliography}
\end{document}